\documentclass[11pt, doublespacing]{article}

\usepackage{fullpage, setspace} 
\usepackage{authblk}
\usepackage{graphicx}
\usepackage{amssymb}
\usepackage{epstopdf}
\usepackage{natbib}
\usepackage{amsmath}
\usepackage{graphics}
\usepackage{bm}
\usepackage{color}
\usepackage{amsthm}
\usepackage{enumerate}
\usepackage{amsfonts}
\usepackage{algorithm}
\usepackage[noend]{algpseudocode}
\usepackage{bbm}

\newcommand{\E}{\mathbb{E}}

\newcommand{\btheta}{\bm{\theta}}
\newcommand{\bthetaz}{\bm{\theta}_0}
\newcommand{\bhtheta}{\hat{\bm{\theta}}_n}
\newcommand{\gbtheta}{g_{\bm{\theta}}}
\newcommand{\gbthetaz}{g_{\bm{\theta}_0}}
\newcommand{\gbhtheta}{g_{\hat{\bm{\theta}}_n}}
\newcommand{\bx}{\bm{x}}
\newcommand{\by}{\bm{y}}
\newcommand{\hdelta}{\hat{\delta}_n}

\newcommand{\innern}[2]{\langle #1, #2 \rangle_n}
\newcommand{\inner}[2]{\langle #1, #2 \rangle}

\newcommand{\bigO}{\ensuremath{\mathop{}\mathopen{}\mathcal{O}\mathopen{}}}
\newcommand{\smallO}{ \scalebox{0.7}{$\mathcal{O}$}}
\newcommand{\bigOp}{\bigO_p}

\newcommand{\supp}[1]{#1}
\newcommand{\revise}[1]{#1}

\newcommand{\ignore}[1]{}

\newtheorem{assumption}{Assumption}
\newtheorem{thm}{Theorem}
\newtheorem{col}{Corollary}

\begin{document}

\title{\vspace*{-1cm}\LARGE\bf A Frequentist Approach to Computer Model Calibration}

\author[1]{Raymond K. W. Wong\thanks{\noindent Department of Statistics, Iowa State University, Ames, IA 50011, U.S.A. Email: {\tt raywong@iastate.edu}}}
\author[2]{Curtis B. Storlie\thanks{\noindent Statistical Sciences Group, Los Alamos National Laboratory, Los Alamos, NM 87545, U.S.A.  Email: {\tt storlie@lanl.gov}}}
\author[3]{Thomas C. M. Lee\thanks{\noindent Department of Statistics, University of California,
    Davis, One Shields Avenue, Davis, CA 95616, U.S.A. Email: {\tt tcmlee@ucdavis.edu}}}
\affil[1]{Department of Statistics, Iowa State University}
\affil[2]{Los Alamos National Laboratory}
\affil[3]{Department of Statistics, University of California, Davis}

\date{September 04, 2015}

\maketitle

\begin{abstract}
This paper considers the computer model calibration problem and provides a general frequentist solution.  Under the proposed framework, the data model is semi-parametric with a nonparametric discrepancy function which accounts for any discrepancy between the physical reality and the computer model.  In an attempt to solve a fundamentally important (but often ignored) identifiability issue between the computer model parameters and the discrepancy function, this paper proposes a new and identifiable parametrization of the calibration problem.  It also develops a two-step procedure for estimating all the relevant quantities under the new parameterization.  This estimation procedure is shown to enjoy excellent rates of convergence and can be straightforwardly implemented with existing software.  For uncertainty quantification, bootstrapping is adopted to construct confidence regions for the quantities of interest.  The practical performance of the proposed methodology is illustrated through simulation examples and an application to a computational fluid dynamics model. 

\end{abstract}
\pagebreak

\section{Introduction}
In many scientific studies, complex mathematical models, implemented as
computer codes, are often used to model the physical reality
\citep[see, e.g.,][]{Santner03, Fang-Li-Sudjianto10}.
Such computer
codes are also known as {\em computer models}, and can only be executed when
certain model parameters are pre-specified.
The goal of computer model {\em calibration} is to find the model parameter values that allow the computer model to best reproduce the physical reality.

In the computer model calibration problem \citep{Kennedy-OHagan01}, an output $y$ is observed from the physical reality $\zeta$ at $n$ locations of a $p$-variate input $\bx=(x_1,\dots,x_p)^{\intercal}$:
\[
y_i=\zeta(\bx_i)+\varepsilon_i, \quad i=1,\ldots,n,
\]
where $\varepsilon_i$ is the measurement error for the $i$-th observation.  It is assumed that the user can select the values of the design locations $\bx_1, \ldots, \bx_n$.  A computer model $\eta(\bx, \btheta)$, also called the {\em simulator}, can be used to approximate the physical reality $\zeta(\bx)$ when its model parameter $\btheta=(\theta_1,\dots,\theta_d)^{\intercal}$ is selected to be close to the unknown ideal value $\btheta_0$.  To account for the discrepancy between the simulator and the physical reality, one can introduce a discrepancy function $\delta_0(\bx)$ and the model of the experimental data $\{\bx_i,y_i\}_{i=1}^{n}$ becomes
\begin{equation}
y_i = \eta(\bx_i, \bthetaz) + \delta_0(\bx_i)  + \varepsilon_i, \quad i=1, \ldots, n.
\label{eqn:data_model}
\end{equation}
To use~(\ref{eqn:data_model}) in practice, one first needs to estimate
$\btheta_0$ and $\delta_0$, which requires evaluation of $\eta(\bx,\btheta)$ at
many different values of $\bx$ and $\btheta$.  However, the evaluation of
$\eta(\bx,\btheta)$ is often very computationally expensive due to the complex nature of the
mathematical models.  This complication can be alleviated via the use of a
surrogate model, often referred to as an {\em emulator}, for the simulator
$\eta(\bx,\btheta)$ \citep[e.g.,][]{Currin-Mitchell-Morris91, Kennedy-OHagan01,
  Higdon-Kennedy-Cavendish04,Drignei-Morris06, Conti-Gosling-Oakley09, Reich-Storlie-Bondell09}.
Typically a Gaussian
process (GP) is assumed for the emulator to allow for a flexible model of
the simulator. Moreover, the discrepancy function $\delta_0$ is also often
modeled by a GP.  In order to construct an emulator, an additional set of
outputs $y'_i$'s generated from the simulator is obtained at $m$ design locations $(\bx'_1,\btheta'_1),\dots,(\bx'_{m},\btheta'_{m})$.  Thus, there are two observed data sets: 
the {\em experimental data} set $\{\bx_i,y_i\}_{i=1}^{n}$ obtained from the physical reality and the {\em simulator data} set $\{\bx'_i,\btheta'_i, y'_i\}_{i=1}^{m}$ generated by the simulator $\eta(\bx,\btheta)$.
\revise{Note that the experimental and simulator data sets are fundamentally different.  The simulator data are generated solely from the simulator and so they are not directly related to the physical reality.  Moreover, the designs of the two data sets do not have to match. That is, $\{\bx_i'\}$ does not necessarily equal to $\{\bx_i\}$ either in value or sampling distribution.}
In this setting, the goal is to estimate $\btheta_0$, $\delta_0$ and $\eta$.  

Traditionally, this estimation problem is solved within a Bayesian framework
\citep[e.g.,][]{Kennedy-OHagan01, Higdon-Kennedy-Cavendish04, Oakley04, Bayarri07, Higdon08, Storlie-Lane-Ryan14}, partly because of the GP's
ability to incorporate uncertainty about the surrogate model for
$\eta(\bx,\btheta)$, and a similar ability to provide uncertainty about the
discrepancy function $\delta_0$.  However, in this paper we are interested in
solving this problem from a purely frequentist perspective with a
  nonparametric discrepancy function, and at the same time also
accounting for uncertainty in the model parameters, the surrogate model, and
the discrepancy function.  To the best of our knowledge, there has been no
previous attempt to solve the computer model calibration problem in this
manner.  Yet, there are several reasons for doing this: 
\begin{enumerate}
\item The proposed approach is conceptually clean and simple, easy to understand, and can be paired with any choice of surrogate model. 
\item It delivers a complementary calibration result to the Bayesian approach, allowing each approach to provide some qualitative and quantitative confirmation of the other. 
\item Researchers using computational models may be opposed to the Bayesian calibration approach because of the complex prior assumptions of GP and of an identifiability issue between the surrogate model and the discrepancy function (see Section~\ref{sec:ident}).
Consequently they may prefer the proposed frequentist approach instead.
\end{enumerate}
\vspace*{-0.3cm}

The proposed frequentist approach explicitly accounts for all potentially
important sources of uncertainty, and is a viable alternative to the Bayesian
approach.  While any statistical model makes assumptions, there are fewer
assumptions necessary in the proposed approach than in the Bayesian
counterpart.  For example, prior distributions on the surrogate model and the
discrepancy function are replaced with ``smoothness'' assumptions and an
emphasis is intuitively placed on the ability to predict the experimental data
via cross-validation.  Empirical evidence shows that the proposed frequentist
approach tends to outperform existing Bayesian methods on the examples
considered.  

Lastly we remark that there are some frequentist solutions for the computer model calibration problem.  The most common one involves obtaining the maximum likelihood estimate (MLE) for $\btheta$ directly by evaluating $\eta$ sequentially in an optimization routine \citep[e.g.,][]{Vecchia-Cooley87,Jones-Schonlau-Welch98,Huang-Allen-Notz06}.  If $\eta$ is computationally expensive then as before a surrogate model could be used in place of $\eta$ for the purpose of obtaining an MLE for $\btheta$.
However, this latter approach must be used with caution as to not ignore the estimation uncertainty in the surrogate model for $\eta$, which can often be substantial.
Also, unlike the proposed approach, most previous frequentist methods
suffer from the same shortcoming: no discrepancy modeling is included in their
formulation (i.e., $\delta_0=0$).
While some surrogate models may be very good
approximations, no model is perfect and neglecting the discrepancy can be a
major pitfall \citep{Bayarri07,Brynjarsdottir14}.
One notable exception is the pioneering work by \citet{Joseph-Melkote09}, which substantially improves previous frequentist model calibration methods by assuming a parametric form for the discrepancy function $\delta_0$.  Our work brings frequentist calibration approaches to to the next level in several respects: (i) we allow for a nonparametric (or semi-parametric) $\delta_0$ to provide a practical representation of ignorance for the discrepancy; (ii) we present a rigorous justification for the proposed estimates, and (iii) we deliver a simple, yet general procedure to account for the uncertainty in the surrogate model, the model parameters, and the discrepancy. 
As discussed in detail below, the use of a nonparametric $\delta_0$ gives rise to an identifiability issue and some theoretical challenges, which separates our work from others.

The rest of this article is organized as follows.  In Section~\ref{sec:semi_calib} we introduce an identifiable \revise{definition} of the calibration problem, and provide \revise{a} practical estimation procedure.  Section~\ref{sec:boot} illustrates how the bootstrap methodology can be applied to provide uncertainty quantification for parameters of interest.  Theoretical backup for our estimation procedure is presented in Section~\ref{sec:theory}.
\revise{The practical performance of our methodology is illustrated via a real life computational fluid dynamics model in Section~\ref{sec:app}, while concluding remarks are given in Section~\ref{sec:conclude}.  Additional material including simulation experiments and technical details are provided in a separate online supplemental document.}


\section{The Proposed Approach}
\label{sec:semi_calib}
Consider the semi-parametric model~(\ref{eqn:data_model}) for the physical reality.  Despite its popularity under the Bayesian framework, this model is not identifiable in the frequentist regime, where $\bthetaz$ and $\delta_0$ are treated as fixed but unknown quantities to be estimated.  In the following we will first discuss this non-identifiability issue and provide intuitive and identifiable definitions for $\bthetaz$ and $\delta_0$ under model~(\ref{eqn:data_model}) (Sections~\ref{sec:ident}).  We then develop an efficient method for estimating these quantities when the simulator $\eta$ is known (Section~\ref{sec:est}) and unknown (Section~\ref{sec:emulator}).  As to be seen below, the proposed frequentist framework is very general and covers many practical situations.  Its estimation procedure can be paired with any existing optimization techniques and nonparametric regression methods, therefore provides an effective and flexible approach to estimate $\bthetaz$ and $\delta_0$ with convenient implementation.

\subsection{An Identifiable Formulation}\label{sec:ident}

In model~(\ref{eqn:data_model}), the discrepancy function $\delta_0$ is assumed to be an unconstrained smooth function, and as such will be estimated using nonparametric regression techniques.
To see the non-identifiability of~(\ref{eqn:data_model}), consider two different values $\bm{\theta}_1$ and $\bm{\theta}_2$ for $\bm{\theta}$, and write 
$\delta_1(\bm{x}) = \zeta(\bm{x}) - \eta(\bm{x}, \bm{\theta}_1)$
and
$\delta_2(\bm{x}) = \zeta(\bm{x}) - \eta(\bm{x}, \bm{\theta}_2)$.
As both $(\btheta_1, \delta_1)$ and $(\btheta_2, \delta_2)$ give the same distribution for $y$, model~(\ref{eqn:data_model}) in general is unidentifiable.  In the Bayesian paradigm, with the help of suitable priors, the posterior distributions for $\bthetaz$ and $\delta_0$ are technically well-defined.
However, this unidentifiability still poses many problems at a more foundational level.  For example, it forbids any meaningful construction of uncertainty measures for $\bthetaz$ or $\delta_0$, as these target quantities do not have unique definitions.  We also note that previous frequentist methods bypass this issue by setting $\delta_0=0$.

Now we provide natural and sensible definitions of $\bthetaz$ and $\delta_0$ to achieve identifiable modeling.  Write the spaces of model parameters and inputs as $\bm{\Theta}$ and $\mathcal{X}$, respectively.
We propose the following model for the physical reality $\zeta$:
\[
  \zeta(\bm{x}) = \eta(\bm{x}, \bthetaz) + \delta_0(\bm{x}),
\]
where $\eta$ is the simulator and $\delta_0$ is the discrepancy function, both to be modeled by smooth functions.  We define 
\begin{equation}
\bthetaz =\underset{\btheta\in\bm{\Theta}}{\operatorname{argmin}}
\int_\mathcal{X} \{\zeta(\bm{x})- \eta(\bm{x},
\btheta)\}^2 dF(\bm{x}),
\label{eqn:thetadef}
\end{equation}
\revise{
where the distribution $F$ characterizes a weighting scheme of $\bm{x}$.
For simplicity, we assume that $F$ is the sampling distribution of $\bm{x}$.
This is mostly sensible since the sampling design, being user controllable, would
characterize one's attention to $\zeta$ over different values of $\bm{x}$.
Thus this should also be reflected in the definition of $\bthetaz$.
For common sampling design schemes, $F$ is intuitively defined as an identity function
and thus $\bthetaz$ is the minimizer of the integrated squared error between
$\zeta$ and $\eta$.
However, if a weighting scheme, other than the sampling distribution of $\bm{x}$, is intended,
our procedure can be modified easily by introducing a simple
reweighting (to $M_n$ defined below in (\ref{eqn:M})) to estimate the
corresponding $\bthetaz$.
}

Under a mild regularity assumption (Assumption~\ref{assumption:iden} below), the solution of the minimization problem in~(\ref{eqn:thetadef}) is unique, and hence it is straightforward to see that both $\bthetaz$ and $\delta_0$ are identifiable.

The definition~(\ref{eqn:thetadef}) for $\bthetaz$ is logical and natural, as it aligns with the intuition that $\bthetaz$ should be the value that makes $\eta$ closest to $\zeta$, and $\delta_0$ is used to account for the remainder.  Using other arguments or motivation, it is of course possible to provide alternative identifiable definitions for $\bthetaz$ and $\delta_0$.  For example, \citet{Brynjarsdottir14} argue that the the ``best fitting'' model is not always the most desirable.  However, our definition~(\ref{eqn:thetadef}) leads directly to a straightforward and easily implementable estimation procedure, as to be described below.  \revise{We also note that our definition is in the same spirit as \citet{Walker13}.}

\subsection{Estimation when the Simulator is Known}
\label{sec:est}
Suppose we observe the output $y$ of the physical reality $\zeta$ at $n$ locations $\bm{x}_1, \dots, \bm{x}_n$; i.e., ${y}_i=\zeta(\bm{x}_i) + \varepsilon_i$, $i=1,\dots, n$, where $\varepsilon_i$ is the $i$-th observation error.  These errors are assumed to be independent and have mean 0.  For simplicity, we assume the $p$-variate $\bm{x}_i$'s have been scaled such that $\mathcal{X}=[0,1]^p$.  With the above modeling of $\zeta$, the observations are assumed to follow
\begin{equation}
  y_i = \eta(\bm{x}_i, \bthetaz) + \delta_0(\bm{x}_i) + \varepsilon_i. \label{eqn:data_model2}
\end{equation}
For the estimation of $\bthetaz$ and $\delta_0$, we first consider a simpler situation for which $\eta$ can be assumed known or evaluated rapidly.  For cases when this assumption is not true, we will estimate $\eta$ with a second set of samples and the details are described later in Section~\ref{sec:emulator}.

The definitions of $\bthetaz$ and $\delta_0$ naturally motivate a two-step estimation procedure:
\begin{enumerate}
  \item {\em Estimation of $\bthetaz$}: compute the estimate $\hat{\bm{\theta}}$ of $\bm{\theta}$ as the solution to the following minimization problem
\begin{equation}
\hat{\bm{\theta}} = \underset{\btheta\in\bm{\Theta}}{\operatorname{argmin}} \ M_n(\btheta) \quad \mbox{with} \quad
 M_n(\btheta)=\frac{1}{n}\sum^n_{i=1}\{y_i - \eta(\bm{x}_i, \btheta)\}^2. \label{eqn:M}
\end{equation}
  \item {\em Estimation of $\delta_0$}: estimate $\delta_0$ by applying any nonparametric regression method to the ``data''
    $\{ \bm{x}_i, y_i-{\eta} (\bm{x}_i, \hat{\bm{\theta}})\}_{i=1}^n$.
\end{enumerate}
This estimation procedure has its beauty in flexibility and ease of implementation.  It can be coupled with any (global) minimization technique in Step~1 and any nonparametric regression method in Step~2.  For example, for convenience one could adopt an existing and fast off-the-shelf minimization routine for Step~1, and wavelet technique for Step~2 if one believes $\delta_0$ is mostly smooth with a few sharp jumps.  Also, as the estimation of $\bthetaz$ and that of $\delta_0$ are separated, there is no need to re-run the minimization in Step~1 when choosing the smoothing parameter in the nonparametric regression step.  Thus in general this estimation procedure can be made very fast computationally with suitable choices of minimization and nonparametric regression techniques.  For all the numerical illustrations in this paper, we adopt the genetic optimization using derivative \citep{Sekhon-Mebane98} for Step~1 and smoothing spline ANOVA \citep{Wahba90} for Step~2.  The theoretical support for using these estimators are provided in Section~\ref{sec:theory} under a fixed design setting.


\subsection{Estimation when the Simulator is Approximated with an Emulator}
\label{sec:emulator}
This subsection handles the situation when the simulator $\eta$ is unknown or expensive to run.  As mentioned in introduction, a common strategy to overcome
this issue is to approximate $\eta$ with a surrogate model, also known as an
emulator.  The emulator is estimated nonparametrically from a second
set of observations obtained by running the simulator for different
combinations of inputs $\bx$ and model parameters $\btheta$.  Write the
simulator output at $m$ design locations
$(\bx'_1,\btheta'_1),\dots,(\bx'_{m},\btheta'_{m})$ as
$\by'=(y'_{1},\dots,y'_{m})^\intercal$.
They are assumed to follow
\[
y'_{j} = \eta(\bx'_j, \btheta'_j) + \tau_j, \quad j=1,\dots,m,
\]
where $\tau_j$'s are independent random errors with mean zero.  The underlying physical model is typically continuous and in theory these $\tau_j$'s should not be needed.  However, they are included here to allow for numerical jitter in the simulator evaluations due to various reasons such as convergence criteria.

The proposed approach proceeds as follows.  We first use
$\{\bx'_j,\btheta'_j,y'_j\}_{j=1}^{m}$ to fit an emulator via a
nonparametric regression method such as SS-ANOVA \citep{Wahba90}.
Denote the resulting emulator as
$\hat{\eta}$.  We then treat $\hat{\eta}$ as fixed and replace $\eta$ by
$\hat{\eta}$ in~(\ref{eqn:data_model2}) when estimating $\bthetaz$ and
$\delta_0$ via the estimation procedure proposed in Section~\ref{sec:est}.
We note that the parameter $\btheta$ can be constrained to a particular domain, as is often done (albeit with the ability to be more informative) in the Bayesian calibration approach via a prior distribution.  

In the above description the estimation of $\eta$ and $\delta_0$ is done separately.  One could alternatively perform a joint estimation by combining the two estimation problems into one semi-parametric optimization problem.  However, in this case (as in the Bayesian approach) the experimental data $\{\bx_i, y_i\}_{i=1}^{n}$ will influence the estimation of the emulator.  This could be beneficial by providing a smaller variance to the emulator, but it could also be problematic by inserting a large bias into the emulator toward the reality function.  The more conservative approach taken here eliminates this potential bias by removing the influence of the experimental data on the emulator.

To the best of our knowledge, this approach to obtain a point estimate for the calibration problem with discrepancy ($\bthetaz$ and $\delta_0$) for a computationally demanding simulator $\eta$ has not been attempted until now.  The above description does not yet account for the uncertainty in the estimation of the emulator, model parameters, or the discrepancy function. However, this issue can easily be addressed via the bootstrap method \citep[see, e.g.,][]{Efron-Tibshirani94}, to be described next.

\section{Uncertainty Quantification using Bootstrap}
\label{sec:boot}

In computer modeling, bootstrap has been applied successfully to quantify the uncertainty in the emulator for the purpose of sensitivity analysis (SA) and uncertainty analysis (UA) for computationally demanding simulators \citep[e.g.,][]{Storlie-Swiler-Helton09, Storlie-Reich-Helton13}.
It is therefore expected that bootstrap will also provide equally successful results for the current calibration problem.  However, it is noted that this calibration problem is far more complicated than SA and UA due to the additional estimation of $\bthetaz$ and $\delta_0$.

Let the point estimates of the unknown parameters in
model~(\ref{eqn:data_model}) be obtained as described above and denoted as
$\hat{\btheta}$, $\hat{\eta}$, and $\hat{\delta}$.  These define an estimate
for the data generating process for both the simulator data $\by'$ and
experimental data $\by$.  A bootstrap sample for the calibration problem can be generated with the following steps:
\begin{enumerate}
  \item (Optional) Re-sample the designs in both data sets if the data were generated under
    random designs.
  \item Produce $B$ bootstrap samples by re-sampling centered residuals.
  \item Re-estimate the parameters to obtain $B$ bootstrap estimates of $\btheta$, $\eta$, and $\delta$.  Denote them as $\hat{\btheta}_b^*$, $\hat{\eta}_b^*$, and $\hat{\delta}_b^*$, $b=1,\dots,B$, respectively.
\end{enumerate}
The resulting bootstrap sample of the estimates can be used to obtain a bootstrap confidence region for most quantities of interest.

In calibration problems, confidence intervals for the elements of $\btheta_0$ and pointwise confidence bands for $\delta_0$ are usually of interest.  For example, to obtain a confidence interval for the first element $\theta_{0,1}$ of $\bthetaz$, one can find the $\alpha/2$ and $(1-\alpha/2)$ sample quantiles from $\{\hat{\theta}_{1,1}^*, \dots, \hat{\theta}_{B,1}^*\}$, where $\hat{\theta}_{b,1}^*$ represents the first element of $\hat{\btheta}^*_b$ for $b=1,\dots,B$.  Denote these quantiles as $z^*_{\alpha/2}$ and $z^*_{1-\alpha/2}$, respectively.  The required confidence interval is then given by 
$(z^*_{\alpha/2} \;,\; z^*_{1-\alpha/2} )$.
A confidence interval for a prediction of the physical reality $\zeta$ at any new input $\bx_{\mbox{\scriptsize new}}$ and the pointwise confidence band for $\delta_0$ can be obtained in a similar fashion.

Since our estimation procedure involves nonparametric regression, the impact of bias may lead to incorrect asymptotic coverage of the aforementioned bootstrap confidence regions \citep[see, e.g.,][]{Hardle-Bowman88, Hall92, Hall92a}.  In the literature, there are two common strategies for correcting the coverage: undersmoothing and oversmoothing.  As shown in \citet{Hall92}, undersmoothing is a simpler and more effective strategy than oversmoothing.  Our estimation procedure can be easily modified to incorporate undersmoothing; e.g., by choosing a smaller smoothing parameter.  However, the gain in practical performance of these strategies are usually small and most of these strategies involve another ad-hoc choice of the amount of under- or over-smoothing.  Moreover, it is not uncommon to ignore this bias issue, essentially resulting in the use of non-adjusted confidence regions as described above; see, e.g., \citet{Efron-Tibshirani94} and \citet{Ruppert-Wand-Carroll03}.  To keep the approach simple and adaptable to a wide class of nonparametric regression methods, we recommend using the non-adjusted confidence regions.

\section{Theoretical Results}\label{sec:theory}
This section provides theoretical support to the proposed estimation procedure presented in Section~\ref{sec:semi_calib}.  First recall that the estimation of $\eta$ depends on a second independent sample generated from the simulator of size $m$.  In practice this sample is typical much larger than the sample obtained from the physical reality; i.e., $m \gg n$.  Thus, it is reasonable to assume that $m$ approaches infinity at a faster rate than $n$ in the asymptotic framework.  If $m \rightarrow \infty$ fast enough, the asymptotics of $\hat{\btheta}$ and $\hat{\delta}$ would be similar to those under known $\eta$.  Therefore, for simplicity and to speed up the development, in the following we assume $\eta$ is known and derive the asymptotic properties of $\hat{\btheta}$ and $\hat{\delta}$ defined in Section~\ref{sec:est}.

Write $\hat{\bm{\theta}}$ and $\hat{\delta}$ as $\bhtheta$ and $\hdelta$
respectively to address their dependence on $n$.  In the following, we assume
that $\bx_1,\dots, \bx_n$ are fixed and
use $F_n$ to denote their empirical distribution function.
In addition, $\|\cdot\|_n$, $\|\cdot\|$ and $\|\cdot\|_E$ represent the
$L_2(F_n)$-norm,
the $L_2(F)$-norm and the Euclidean norm respectively.
For two functions $g$ and $h$, let $\innern{g}{h}=\sum^n_{i=1}g(\bx_i)h(\bx_i)$
and $\inner{g}{h}=\int_{\mathcal{X}}g(\bx)h(\bx)dF(\bx)$.  With slight notation
abuse, we also write $\innern{y}{g}=(1/n)\sum^n_{i=1}
y_ig(\bx_i)$ and $\innern{\varepsilon}{g}=(1/n)\sum^n_{i=1} \varepsilon_i
g(\bx_i)$.  Lastly, write ${\gbtheta}(\bx) = \eta(\bx,\btheta)$,
$\mathcal{G}=\{\gbtheta: \btheta\in\Theta\}$ and $\mathcal{G}-g =
\{\gbtheta-g:\btheta\in\Theta\}$ for a function $g$.
\revise{And, use $g^{(j)}_{\btheta}$ to represent the $j$-th order derivative of $g_{\btheta}$ with respect to $\btheta$ for $j=1,2$.}

When deriving asymptotic results for similar statistical problems, it is relatively common to assume an independent and identically distributed (i.i.d.) random design, as it is easier than a fixed design to work with.  However, for most practical calibration problems, the design is either fixed or correlated (e.g., Latin Hypercube sampling).  Therefore the following results are developed under a fixed design, despite it is a more challenging setting than the i.i.d.~random design.  Note that model~(\ref{eqn:data_model2}) is a semi-parametric model and we first approach the parametric part and establish the $\sqrt{n}$-consistency of $\bhtheta$ in (see Theorem~\ref{thm:theta}), where the difficulty lies in the existence of the discrepancy function $\delta_0$.  The effect is similar to a regression model with misspecification.

As for the nonparametric part, $\delta_0$, we adopt the framework of Section 10.1 of \citet{Van-De-Geer00} for penalized least squares estimation.  We extend Theorem~10.2 of \citet{Van-De-Geer00} to obtain the asymptotic behavior of $\hdelta$ (see
 \supp{Lemma 2 of the supplemental document}
  and Theorem \ref{thm:delta:general}), taking into account the effect of estimation error of $\bhtheta$.  Let the class of functions to which $\delta_0$ belongs be $\mathcal{H}$. We suppose that $\mathcal{H}$ is a cone.  Under van de Geer's framework, the general form of the estimate of $\delta_0$ is
\begin{equation}
  \hdelta = 
\underset{\delta\in\mathcal{H}}{\operatorname{argmin}} 
\left[\frac{1}{n} \sum^n_{i=1}
    \{y_i - \gbhtheta(\bx_i) - \delta(\bx_i)\}^2 + \lambda_n^2 J^{v}(\delta) \right],
  \label{eqn:delta}
\end{equation}
where $v>0$, $\lambda_n>0$, $J:\mathcal{H}\rightarrow [0,\infty)$ is a
pseudo-norm on $\mathcal{H}$. The $\lambda_n$ is known as the smoothing parameter.

As an illustration, we provide the convergence rate of $\hdelta$ for $p=1$ if a penalized smoothing spline is used (see Corollary \ref{col:smoothing}).  This requires an additional orthogonality argument for the application of Theorem~\ref{thm:delta:general}.  We will write $\bx$ as $x$ when $p=1$.

Below are the assumptions needed for our theoretical results.

\begin{assumption}[Error structure]\label{assumption:err}
$\E(\varepsilon_i)=0$ {and $\E(\varepsilon^2_i)=\sigma^2$} for $i=1,\dots,n$. Also,
$\varepsilon_1,\dots, \varepsilon_n$ are uniformly sub-Gaussian; that is, there {exist} $K$
and $\sigma_0$ such that
\[
  \max_{i=1,\dots,n} K^2 \left\{\E \exp(\varepsilon_i^2 /K^2) - 1\right\} \le
  \sigma_0^2.
\]
\end{assumption}

\begin{assumption}[Parameter space]\label{assumption:param}
  $\Theta$ is a totally bounded $d$-dimensional Euclidean space. That is, 
  there exists {an} $R_0>0$ such that \revise{$\|\btheta\|_E\le R_0$ for all $\btheta\in\Theta$}.
\end{assumption}

\begin{assumption}[Function class $\mathcal{G}$]\label{assumption:g} $ $
  \begin{enumerate}[(a)]
    \item There exists {a} $c_0>0$ such that
      $\|g_{\btheta} - g_{\btheta'}\|_n \le c_0 \| \btheta - \btheta'\|_E$
      for all $\btheta, \btheta'\in\Theta$.
    \item $\gbtheta$ is  twice continuously differentiable with respect to
      $\btheta$ in a neighborhood of $\bthetaz$.
      $\gbtheta^{(1)}(\bx)$ and $\gbtheta^{(2)}(\bx)$ are continuous with
      respect to $\bx$ over this neighborhood.
    \item $\sup_{\bx\in\mathcal{X}}|\gbtheta^{(1)}(\bx)|$ and
      $\sup_{\bx\in\mathcal{X}}|\gbtheta^{(2)}(\bx)|$ are bounded uniformly over a neighborhood of $\bthetaz$.
  \end{enumerate}
\end{assumption}

\begin{assumption}[Convergence of design]\label{assumption:design} $ $
  \begin{enumerate}[(a)]
    \item $\sup_{\btheta\in\Theta}|\|\zeta - \gbtheta\|_n^2 - \| \zeta - \gbtheta\|^2|
    =\smallO(1)$.
    \item \revise{Elements of $A_n-A$ are $\smallO(1)$, where
    \[
    A_n=(1/n)\sum^n_{i=1} \{\gbthetaz^{(1)}(\bx_i)\gbthetaz^{(1)}(\bx_i)^\intercal
    - \delta_0(\bx_i)\gbthetaz^{(2)}(\bx_i)\},
    \]
    \[
    A=\int_\mathcal{X} \gbthetaz^{(1)}(\bx)\gbthetaz^{(1)}(\bx)^\intercal - \delta_0(\bx)\gbthetaz^{(2)}(\bx) dF(\bx),
    \]
    are the second derivative of $M_n(\btheta)$ evaluated at $\bthetaz$ and that
    of $M(\btheta)$ respectively.}
      \item \revise{Euclidean norm of the first derivative of $M_n$ evaluated at $\bthetaz$, $(1/n)\sum^n_{i=1} \delta_0(\bx_i) \gbthetaz^{(1)}(\bx_i)$,
  is $\bigO(n^{-1/2})$.}
  \end{enumerate}
\end{assumption}

\begin{assumption}[Identification]\label{assumption:iden}
\revise{$A$ is strictly positive definite.}
\end{assumption}

\begin{assumption}[Discrepancy function]\label{assumption:delta} $ $
  \begin{enumerate}[(a)]
    \item $\delta_0$ is \revise{bounded.}
    \item \revise{$J(\delta_0)<\infty$.
    \item There exist $K_1>0$ and $0<\alpha<2$ such that
    $H(u,\tilde{\mathcal{H}} , F_n) \le
  K_1u^{-\alpha}$,
    for all $u>0$ and $n \ge 1$,
    where 
      \[\tilde{\mathcal{H}}=\left\{\frac{\delta - \delta_0}{J(\delta) + J(\delta_0)}:
      \delta\in\mathcal{H}, J(\delta) + J(\delta_0)>0 )\right\}\]
   and $H(u,\tilde{\mathcal{H}}, F_n)$
   is the $u$-entropy of $\tilde{\mathcal{H}}$ for the $L_2(F_n)$-metric \citep[Definition 2.1 of][]{Van-De-Geer00}}.
    \item \revise{$\sup_{\gamma\in\tilde{\mathcal{H}}}\|\gamma\|_n < \infty$.}
  \end{enumerate}
\end{assumption}

The two main theorems and a corollary now follow.  \supp{Their proofs can be found in Section S3 of the supplemental document.}

\begin{thm}[Rates of convergence of $\bhtheta$ and $\gbhtheta$]
  \label{thm:theta}
\revise{Suppose} that Assumptions
\revise{\ref{assumption:err}-\ref{assumption:iden} and
\ref{assumption:delta}(a)}
hold. We have $\|\bhtheta- \bthetaz\|_E=\bigOp(n^{-1/2})$ and
$\|\gbhtheta-\gbthetaz\|_n=\bigOp(n^{-1/2})$.
\end{thm}


\begin{thm}[Rate of convergence of $\hdelta$]\label{thm:delta:general}
  \revise{Suppose that Assumptions \ref{assumption:err}-\ref{assumption:delta}
hold. Moreover, assume that $v>(2\alpha)/(2+\alpha)$.
\begin{enumerate}[(i)]
  \item
    If $J(\delta_0)>0$ and $\lambda_n \asymp n^{-1/(2+\alpha)}$,
    we have 
      $\|\hdelta-\delta_0\|_n = \bigOp(n^{-1/(2 + \alpha)})$.
  \item If $J(\delta_0)=0$ and $J(\delta)>0$ for all $\delta\in\mathcal{H}\setminus\{\delta_0\}$,
    we have
    \[
      \|\hdelta -\delta_0\|_n = \bigOp\left(\max\left\{n^{-1/2},
        \lambda_n^{-2\alpha/(2v-2\alpha+v\alpha)} n^{-v/(2v-2\alpha+v\alpha)}\right\}\right).
    \]
  \end{enumerate}
}
\end{thm}


\begin{col}[Penalized smoothing spline]\label{col:smoothing}
\revise{Assume $p=1$, $m\in\{1,2,\dots\}$, $\mathcal{H}=\{\delta:[0,1]\rightarrow \mathbb{R},
\int^1_0 \{\delta^{(m)}(x)\}^2 dx <\infty\}$ and
$J(\delta) = [\int^1_0 \{\delta^{(m)}(x)\}^2 dx]^{1/2}$.
And $\hdelta$ is given in (\ref{eqn:delta}) with $v=2$.
Suppose the conditions of Theorem \ref{thm:theta} hold and $J(\delta_0)<\infty$.
Let $\bm{\psi}=(\psi_1,\dots,\psi_m)^\intercal$ where $\psi_k(x)=x^{k-1}$ for $k=1,\dots,m$.
Further, assume that the smallest eigenvalue of $\int\bm{\psi}\bm{\psi}^\intercal dF_n$ is bounded away
from 0.
\begin{enumerate}[(i)]
  \item
    If $J(\delta_0)>0$, $\lambda_n \asymp n^{-m/(2m+1)}$,
    we have
        $\|\hdelta-\delta_0\|_n = \bigOp(n^{-m/(2m + 1)})$.
  \item If $J(\delta_0)=0$,
    we have
    \[
      \|\hdelta -\delta_0\|_n = \bigOp\left(n^{-1/2}\max\left\{1,
        \lambda_n^{-1/(2m)} \right\}\right).
    \]
  \end{enumerate}}
\end{col}

\section{Application to a Computational Fluid Dynamics Model}
\label{sec:app}
In this section, the proposed approach is applied to a computational fluid dynamics (CFD) model of a bubbling fluidized bed \citep{Lane13}.  The experimental apparatus used to produce the field data used here is the the Carbon Capture Unit (C2U) housed at National Energy Technology Laboratory.  The C2U unit is a bench-top carbon capture system, designed to mimic a post-combustion capture device that could be applied at a coal fired power plant.
\supp{Section S1 of the supplemental document}
 provides an illustration of the C2U system.  A gas mixture (i.e., flue gas) flows through the bottom of the adsorber pictured on the right of \supp{the figure shown in
 Section S1 of the supplemental document}
   and into the bed of solid sorbent, resulting in fluidization of the sorbent.  At low temperature ($\sim40^\circ\:$C), CO$_2$ will chemically bond to the solid sorbent and be effectively lifted out of the gas mixture.  The solid sorbent would then circulate out of the adsorber, be stripped of CO$_2$, and flow back into the adsorber.  However, in this example, the goal was to isolate the fluid dynamics of the bubbling fluidized bed, and thus there is only nitrogen gas flowing through the solid sorbent material in the bed (i.e., no CO$_2$ adsorption is taking place).  These data were collected as part of Department of Energy's (DOE's) Carbon Capture Simulation Initiative (CCSI) \citep{Miller13}.  The open source CFD code Multiphase Flow with Interphase eXchanges (MFIX) \citep{Benyahia12} was used as the simulator of the bubbling fluidized bed.  The experimental setup and the MFIX model used to simulate it are fully documented in \cite{Lai14}.  Below, only an abridged description of the data is provided.

The variables involved in the experimental data are the input variables flow rate ($x_1$, {\tt FRate}) and bed temperature ($x_2$, {\tt Temp}), and the output variable ($y$) is the pressure drop at location P3820 (i.e., the pressure drop across the bubbling fluidized bed).  The pressure drop output is the time averaged value of the pressure drop once it was oscillating in steady state. The P3820 pressure drop was measured on the physical C2U system at a design of 44 distinct input settings.  A total of 60 MFIX simulation cases were also designed and run for the purpose of emulator estimation.  Both data sets are available online at the journal website.  The MFIX model parameters involved in the calibration for this case were Res-PP ($\theta_1$): the particle-particle coefficient of restitution, Res-PW ($\theta_2$): the particle-wall coefficient of restitution, FricAng-PP ($\theta_3$): the particle-particle friction angle, FricAng-PW ($\theta_4$): the particle-wall friction angle, PBVF ($\theta_5$): Packed bed void fraction, and Part-Size ($\theta_6$): the effective particle diameter of the sorbent material.  The allowable ranges of the model parameters were chosen to be the same as those devised in \cite{Lai14}, mostly from literature review.  

Table~\ref{tab:CFD} provides the estimated $\btheta$ values along with 95\% confidence intervals and credible intervals, respectively, for the proposed \textsf{fboot-rs} method and the \textsf{bss-anova} method, \supp{both are numerically tested by simulations described in Section S2 of the supplemental document:}
\revise{
\begin{enumerate}
  \item \textsf{fboot-rs}: The proposed frequentist method coupled with the bootstrap procedure of Section~\ref{sec:boot} {\em with} re-sampling of the design (i.e., keep Step~1).
  \item \textsf{bss-anova}: Calibration of computational models via Bayesian
    smoothing spline ANOVA \citep{Storlie-Lane-Ryan14}.
\end{enumerate}
}
Both methods largely agree on their respective estimates and CIs, which provides some confirmation of the result.  The first five parameters have fairly wide CIs relative to their allowable ranges, indicating that most of the range of these parameters produces reasonable model results.  However $\theta_6$ (effective particle size) does have tighter CIs and it seems as though values closer to 117 are preferred.

\begin{table}
  \caption{\label{tab:CFD}Estimates of $\btheta$ along with 95\% Confidence Intervals and Credible Intervals, respectively, for the proposed \textsf{fboot-rs} method and \textsf{bss-anova}.  The parameters were restricted the the ranges provided during the estimation procedures.}
  \centering
  \fbox{
    \begin{tabular}{c|c|cc|cc}
\multicolumn{2}{c}{} &\multicolumn{2}{|c|}{\textsf{fboot-rs}} & \multicolumn{2}{c}{\textsf{bss-anova}} \\
\hline
      $\bm{\theta}_0$ & Range & $\hat{\bm{\theta}}$ & 95\% CI & $\hat{\bm{\theta}}$ & 95\% CI \\\hline
      $\theta_{0,1}$ & [0.80, 1.00] & 0.927& (0.828,  0.964) & 0.908 &  (0.831, 0.978)\\
      $\theta_{0,2}$ & [0.80, 1.00] & 0.831& (0.829,  0.969) & 0.897 &  (0.823, 0.979)\\
      $\theta_{0,3}$ & [25, 45] & 39.5& (26.1,  41.5) & 31.1 & (25.4, 39.8)\\
      $\theta_{0,4}$ & [25, 45] & 33.4& (25.5,  39.3) &  31.4 & (25.3,41.0)\\
      $\theta_{0,5}$ & [0.30, 0.40] & 0.346& (0.316,   0.388) &   0.349 & (0.313, 0.386)\\
      $\theta_{0,6}$ & [99, 125] & 117& (110, 120) & 115 & (108, 120)\\
    \end{tabular}
  }
\end{table}

\begin{figure}
\centering
\makebox{\includegraphics[width=.9\textwidth]{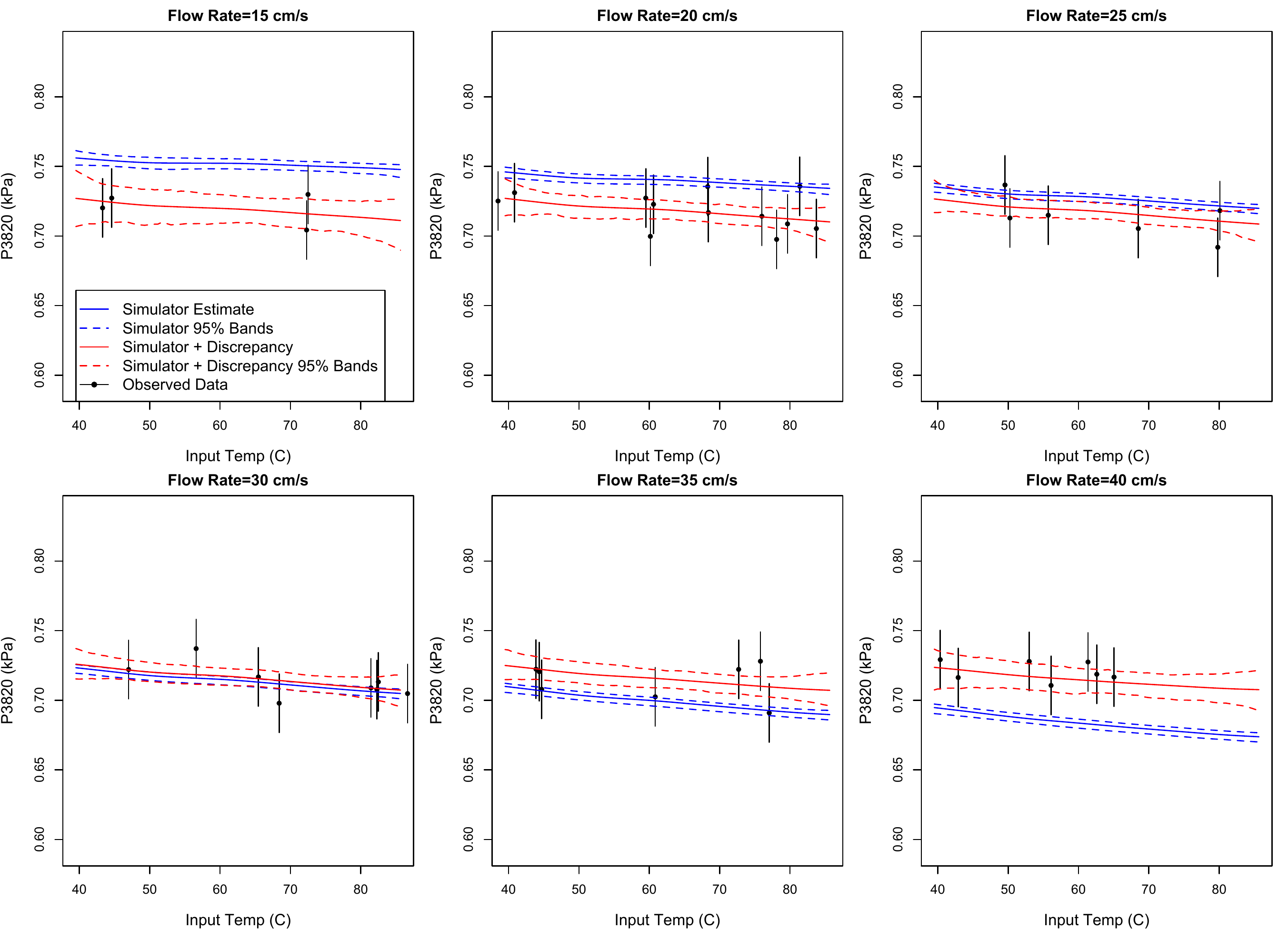}}
  \caption{Fitted simulator (emulator) and simulator plus discrepancy curves (along with 95\% confidence bands) as a function of Temperature at six distinct Flow Rates.  Experimental data is binned into the closest value of the displayed Flow Rates for display.}\label{fig:CFD_fit}
\end{figure}

\begin{figure}
\centering
\makebox{\includegraphics[width=.7\textwidth]{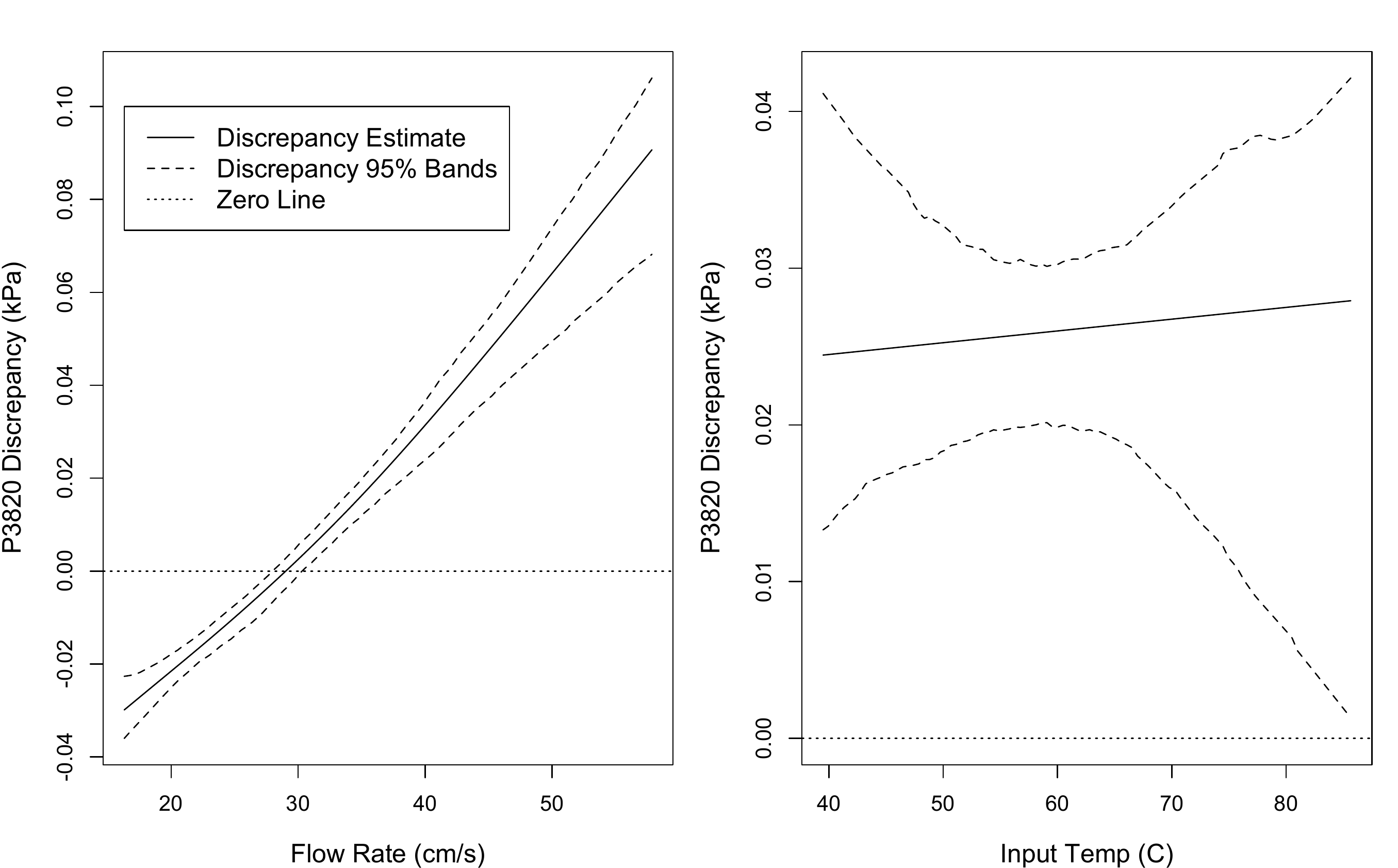}}
  \caption{P3820 Discrepancy main effects across Flow Rate and Temperature.}\label{fig:CFD_dis}
\end{figure}

Figure~\ref{fig:CFD_fit} provides a visual summary for the simulator fit to the experimental data along with confidence bands (accounting for uncertainty in the emulator and the value of $\btheta$).  The simulator with discrepancy (i.e., reality) predictions are provided as well.  The pressure drop ($y$) is plotted against {\tt Temp} ($x_2$) for six distinct values of {\tt FRate} ($x_1$).  The experimental data is also provided (along with the estimated 2$\sigma$ measurement error bars).  The experimental data was binned into the closest value of the six displayed {\tt FRate} for display purposes.  It is clear that the discrepancy is trending upwards as {\tt FRate} increases.  Figure~\ref{fig:CFD_dis} makes this relationship more explicit by isolating the discrepancy main effect functions across {\tt FRate} and {\tt Temp}, respectively.  While it is evident that there is some statistically significant model discrepancy here, such discrepancy is relatively small when considering the magnitude of the pressure drop: the discrepancy is on the order of $\sim$ 0.05 kPa, while the pressure drop is on the order of 0.72 kPa, thus a relative error of roughly 7\%.  Thus, for practical purposes MFIX can be used for prediction of a bubbling fluidized bed, knowing that the model form discrepancy is negligible.

\section{Concluding Remarks}
\label{sec:conclude}
In this work, we have provided a frequentist framework for computer model calibration.  This framework applies a general semi-parametric data model with an emulator for expensive simulators and a discrepancy function, which allows discrepancy between the simulator and the physical reality.  Despite the flexibility of the model, our proposed framework gives identifiable parametrizations for both the model parameters and the discrepancy function.  
A practical estimation procedure and theoretical guarantees are provided for the proposed framework.  Finally, a bootstrapping approach to provide uncertainty quantification for virtually any quantity of interest has also been developed.  Due to the simplicity of the proposed calibration framework and the corresponding bootstrap, this approach can be easily coupled with a variety of optimization methods and/or emulators, which is beneficial to practitioners.

\revise{
Three objectives of calibration are identified by \citet{Brynjarsdottir14}: to study the true values of the physical parameters, to predict the physical system's behaviour within the scope of the observed data (i.e., interpolation) and outside the scope of the data (i.e., extrapolation).  It is important to examine the usefulness of the proposed approach with respect to these objectives.

The proposed approach was not designed to, in general, address the first objective, unless the discrepancy $\delta$ is negligible and $\btheta$ represents the physical parameters.  The proposed approach, however, should work well for the second interpolation objective, although at times it may not be clear what is the phyiscal meaning of $\btheta$.  The third objective (extrapolation) is much more challenging.  As pointed out by \citet{Brynjarsdottir14}, the key to successful extrapolation requires that both $\btheta$ and $\delta$ be meaningfully defined, and that accurate prior information for them are available.  Under these situations Bayesian approach would be useful, and it is not advisable to use the proposed approach, especially when $\delta$ is large.  

This paper is not suggesting that the proposed approach is correct for all problems.  However, when there is no good prior information about some of the model parameters and/or discrepancy, other approaches may not be applicable or could produce unreliable results.  It is in such cases that the proposed approach can be beneficial above and beyond the general benefits provided by the analytic study of its properties.
}


{
\bibliographystyle{rss}
\bibliography{raywongref}
}


\end{document}